\documentclass[11pt,a4paper]{article}  
\usepackage{a4wide}
\begin{document}

\title{Bootstrap equations for effective theories 
and the calculation \\ of the
$G_T/G_V$ ratio}
\author{A. Vereshagin \\
{\em University of Bergen, Department of Physics, Bergen, Norway.}}
\date{November 30, 2001}
\maketitle

\begin{abstract}
A method is described for dealing with effective theories of hadron scattering. 
It allows one to reduce the number of independent renormalization prescriptions 
in those theories and gives a possibility to make numerical predictions. As an 
illustration, we show the results of comparison with the known data on
$\pi\pi$, $\pi K$ and $\pi N$ elastic scattering. This work presents a
generalization and the further development of our results first discussed at the
MENU'99 Symposium
\cite{MENU99}.
\end{abstract}

\section{Preliminary notes}

It is widely believed that to construct the complete theory of
strong processes one needs to make two steps:
\begin{enumerate}
\item With the help of QCD (which is supposed to be the fundamental theory of 
strong forces) find the hadronic spectrum (poles of the Green functions) and 
construct the complete set of asymptotic states.
\item Construct the scattering theory for those (composite) states.
\end{enumerate}

It is not yet clear how to solve the first problem. Is it possible
to say anything about the scattering theory of hadrons, having no
information on their inner structure? That is how the
{\em effective theory}\footnote{
Here this term is understood precisely in the same sense as in
\cite{Weinb2}.} 
concept naturally comes to mind.

When constructing a theory of hadron scattering, we are forced to rely 
upon the Dyson series for the 
$S$-matrix, because this is the only known perturbative approach guaranteeing 
unitarity, causality and Lorentz-invariance of the results (see, e.g.,
\cite{Weinb1}). In the case of effective theory this series can always be 
presented in the form
\begin{equation}
S_{fi}=\langle f | T_{\stackrel{\:}{\scriptscriptstyle\!\! W}}
\exp\left\{ -i \int H_{\rm int} dx \right\} | i \rangle \, ,
\label{2}
\end{equation}
where
$T_{\stackrel{\:}{\scriptscriptstyle\!\! W}}$ stands for Wick's (explicitly 
covariant) T-product and the Hamiltonian density (in the interaction picture)
$H_{\rm int}$ does not contain any noncovariant terms. Imposing any 
algebraic\footnote{Linearly realized, for example isotopic 
$SU_2$ (or $SU_3$).} 
symmetry requirements on a theory based on the form
(\ref{2}) looks not more difficult than if the Lagrangian picture is used: the
$S$-matrix is covariant and satisfies the symmetry requirements, if
$H_{\rm int}$ does.

In contrast, the problem of
{\em dynamical} symmetries looks much more transparent in the Lagrangian 
picture. And the transition from the Lagrangian picture to the Hamiltonian one 
looks almost hopeless when one deals with an effective theory. Indeed, in this 
case one needs to solve an infinite system of constraints arising, in 
particular, due to the presence of higher powers of time derivatives. However, 
as shown in
\cite{Weinb2}, one needs only a
{\em finite} number of Lagrangian terms when working to a given order in a
{\em small momentum}. Thus, in the last case one can construct the corresponding 
Hamiltonian and, hence, avoid problems with unitarity. This program (first 
realized in
\cite{Gasser1}) gives us the natural way 
({\em ``matching''}) to take into account the dynamical symmetry requirements in 
the effective theory based on the Hamiltonian. Namely, one needs to compute the 
amplitude of the process in question, expand it in powers of a small momentum 
(of the Goldstone particle) up to a given order, and, finally, compare the 
result with that following from the canonical approach (based on the invariant 
Lagrangian of the same order) and equate the corresponding constants.  
 
Which fields should be included in the Hamiltonian? To be able to work 
{\em not} only in the low energy region, we include the
{\em resonance fields} (like the
$\rho$-meson) as well as the fields of the true asymptotic 
states (stable with respect to the strong forces, like the
$\pi$-meson).\footnote{
No problem with unitarity occurs in spite of the fact that Dyson's series is 
based on the Hamiltonian depending on resonance fields (see, e.g.
\cite{Veltman}). In fact, this approach is used in the Standard Model of
electroweak interactions: for instance, the
$W$-boson is not an asymptotic state.}
To avoid model dependence of the results, we reserve the possibility to work 
with an arbitrary (possibly infinite) number of resonances with arbitrarily high 
values of spin 
$J$ and mass
$M$. The only limitation is suggested by experiment: we imply that there is a 
finite number of resonances with the same mass (though the mass spectrum may be
unbounded). To put it another way, we imply the existence of a leading Regge 
trajectory (in the
{\em real} plane of
${\rm Re}M$ and 
$J$) which, however, is not necessarily linear. According to the phenomenology 
of 
strong interactions we do not deal with massless spin
$J>{1\over 2}$ particles. Also, we assume that the maximal isospin value is
$I=1$ for mesons and
$I=\frac{3}{2}$ for baryons.

Thus, we consider the effective hadron scattering theory based on Dyson's series 
(\ref{2}) with the Hamiltonian written in the form of an infinite sum of 
Lorentz-invariant (and SU(2)-invariant) local terms, each one constructed 
from the fields (and all powers of their derivatives of arbitrary high order) of 
pions, 
$K$-mesons, nucleons and all possible resonances.

\section{Essential parametres, self-consistency and the bootstrap equations}

It is possible to show that the 
{\em essential 
parametres}\footnote{These are the only parametres that appear in the
$S$-matrix elements, see \cite{Weinb1}, Chapter 7.} 
of the effective scattering theory are masses (real parts of pole 
positions) and those (and only those) combinations of coupling constants which 
are needed to fix the
{\em on-shell} kinematic structure of tree-level vertices. When computing the 
$S$-matrix elements one does not need to impose a renormalization condition on 
each coupling constant appearing in the Hamiltonian: only the essential 
parametres require fixing of their finite parts.

The central idea of our work is that one cannot take independent renormalization 
prescriptions even for the essential parametres of the effective scattering 
theory: certain natural self-consistency requirements impose an infinite number 
of constraints (bootstrap conditions) on the allowed
{\em physical} values of the essential parametres. Namely, to make it possible 
to construct the one-loop approximation for the amplitude of a given scattering 
process (here we only discuss 
$2 \rightarrow 2$ processes), the corresponding
{\em tree-amplitude}
$A(s,t,u)$ must satisfy the following two 
requirements\footnote{Both of them are trivial if there is a finite 
number of terms in the Hamiltonian (Lagrangian).}:
\begin{enumerate}
\item It must be a
{\em meromorphic 
function}\footnote{No singularities except poles.}
of the Mandelstam variables 
$s,t,u$, with poles and residues fixed by the Feynman rules.
\item This amplitude must be
{\em polynomially bounded}\footnote{Polynomial boundedness of the meromorphic 
functions is understood as in complex analysis, see e.g.
\cite{Complex}.}
in each independent energy-like variable at zero value of the corresponding 
momentum transfer.
\end{enumerate}
As explained in
\cite{AVVVVV}, these two requirements turn out to be sufficient to derive the
exact form of the tree-amplitude. At the same time they lead to an infinite 
set of equations connecting the tree-amplitude parametres among themselves
({\em bootstrap equations}). And if we write the Hamiltonian in terms of 
physical parametres (plus the necessary counterterms --- what we can always 
do), then the tree-amplitude is automatically written in terms of 
{\em physical} (experimentally measurable) parametres.
All this means that 
{\bf the bootstrap equations are not affected by the renormalization procedure 
and can be tested experimentally}. 

\section{Comparison with experiment}

In the cases of
$\pi \pi$ and $\pi K$ elastic scattering (see
\cite{AVVVVV,VVV}) it has been found that the resulting equations strongly 
contradict the known data unless two light scalar resonances are taken into 
account. These are the
$\sigma$ ($0^+ 0^+$) and
$\kappa$ ($0^+ \frac{1}{2}^+$) mesons with the following parametres estimated 
from the bootstrap equations: 
$m_\sigma\sim$ 500 MeV; 
$\Gamma_\sigma\sim$ 300 MeV; 
$m_\kappa\sim$ 1 GeV; 
$\Gamma_\kappa\sim$ 500 MeV. 

These parametres are strongly supported by modern data, see, e.g. 
\cite{AVVVVV,UFN}. It is interesting to note, that, as was shown in
\cite{Schechter} (see also 
\cite{UFN}), to preserve the unitarity bound for the 
$\pi\pi$ and 
$\pi K$ amplitudes one must take into account both the resonance and the 
(automatically implied in our approach) background interaction terms.

Perhaps, the most interesting result has been obtained from the analysis of the 
bootstrap equations for the 
$\pi N$ scattering amplitude parametres. It was possible to make the accurate 
estimate of the ratio
$\left. G^{T}_{NN\rho} \right/ G^{V}_{NN\rho} = 6 (\pm 20\%)$, of tensor/vector
$NN\rho$ coupling
constants\footnote{For the corresponding experimental data and notations see
\cite{Nagels}.}. This value turned out to be in nice agreement with experimental 
data. As far as we know, such a relation has never been explained in terms of 
model-independent theoretical arguments.

Besides, with the help of the bootstrap equations we have estimated the values 
of 40 coefficients in the expansion of the
$\pi N$ amplitude around the crossing symmetry point, first introduced in
\cite{Hohler}. The detailed analysis will be published elsewhere.

\subsection*{Acknowledgments}

I wish to thank V.~Cheianov, P.~Osland, F.~Sannino, J.~Schechter,
V.~V.~Sukhanov, A.~N.~Vassilev and especially V.~Vereshagin for stimulating
discussions. This work was supported in part by RFBR (grant 01-02-17152), INTAS 
(INTAS call 2000, project 587), and by Russian Ministry of Education (programme 
``Universities of Russia'' and grant E00-3.3-208).


\end{document}